\documentclass[aps,prl,twocolumn,superscriptaddress,showpacs,floatfix]{revtex4}

\usepackage{graphicx}

\begin{document}

\title{Attractive vortex interaction and the
  intermediate-mixed state of superconductors}

\author{Ernst Helmut Brandt}
\affiliation{Max-Planck-Institut f\"ur Metallforschung,
   D-70506 Stuttgart, Germany}

\author{Mukunda P. Das}
\affiliation{Deptartment of Theoretical Physics, The Australian
   National University, Canberra, ACT 0200, Australia}

\date{\today}

\begin{abstract}
 The magnetic vortices in superconductors usually repel each
 other. Several cases are discussed when the vortex interaction
 has an attractive tail and thus a minimum, leading to vortex
 clusters and chains. Decoration pictures then typically look like
 in the intermediate state of type-I superconductors, showing
 lamellae or islands of Meissner state or surrounded by Meissner
 state, but with the normal regions filled with Abrikosov vortices
 that are typical for type-II superconductors in the mixed state.
 Such intermediate-mixed state was observed and investigated in
 detail in pure Nb, TaN and other materials 40 years ago;
 last year it was possibly also observed  in MgB$_2$, where it
 was called ``a totally new state'' and ascribed to the existence
 of two superconducting electron bands, one of type-I and one
 of type-II. The complicated electronic structure of MgB$_2$ and its
 consequences for superconductivity and vortices are discussed.
 It is shown that for the real superconductor MgB$_2$
 which possesses a single transition temperature, the assumption
 of two independent order parameters with separate penetration
 depths and separate coherence lengths is unphysical.

\end{abstract}

%\pacs{74.25.Qt, 74.25.Sv}

\maketitle

\section{1. Vortex Lattice from GL Theory}

  The most successful phenomenological theory of superconductors
was conceived in 1950 by Ginzburg and Landau (GL) \cite{1}. When
written in reduced units (length unit $\lambda$, magnetic field
unit $\sqrt2 H_c$, energy density unit $\mu_0 H_c^2$, where
$\lambda$ is the magnetic penetration depth and $H_c$ the
thermodynamic critical field) the GL theory contains only {\em one
parameter}, the GL parameter $\kappa = \lambda /\xi$ ($\xi$ is the
superconducting coherence length). The kind of solutions of GL
theory is quite different when $\kappa <1/\sqrt 2$ or $\kappa
>1/\sqrt 2$. Physically, this is due to the fact that the energy
of the wall between a normal and a superconducting domain is
positive when $\kappa < 0.71$ (type-I) and negative when $\kappa >
0.71$ (type-II). This implies that a type-II superconductor is
unstable against the spontaneous formation of many small
superconducting and normal domains. A thesis student of Lev Landau
in Moscow, Alexei Abrikosov in 1953 discovered a periodic solution
of the GL theory which describes the occurrence of a regular
lattice of vortex lines (published in 1957 \cite{2}). Each
Abrikosov vortex (or fluxon, flux line) carries one quantum of
magnetic flux $\Phi_0 = h/2e = 2.07\cdot 10^{-15}$ Tm$^2$, and the
supercurrent circulates around a singular line on which the
complex GL function $\psi(x,y,z)$ (or superconducting order
parameter $|\psi|^2$) is zero. For finding this vortex lattice
solution, in which Landau initially did not believe \cite{2},
Abrikosov obtained the Nobel prize in physics 50 years later in
2003.
 \begin{figure}[htb]   % 1 %%%%%%%%%%%%%%%%%%%%%%%%%%%%%%%%%%%%%%
 \centering
 \includegraphics[scale=.42]{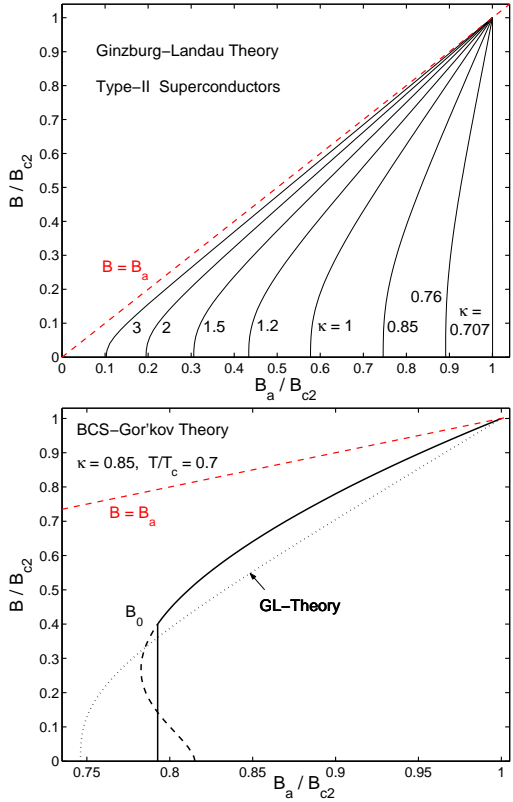}    % .39 .45
\caption{\label{fig1}
{\it Top:~} Ideal reversible magnetization curves of a type-II
superconductor with no vortex pinning and no demagnetizing effects
(long cylinder or slab in a parallel magnetic field $B_a$) from
Ginzburg-Landau Theory. Plotted is the induction $B$ versus $B_a$
for several GL parameters $\kappa = 1/\sqrt2~...~3$.
{\it Bottom:~} The same magnetization curve $B(B_a)$ from the
microscopic BCS-Gor'kov theory for GL parameter
$\kappa \approx 0.85$ and temperature $T/T_c \approx 0.7$,
schematic from \cite{28}. The dashed part for $B < B_0$ is
unphysical, the real $B(B_a)$ contains the depicted vertical line
from $B=0$ to $B=B_0$ obtained from a Maxwell construction that
divides the area under the dashed curve into equal halves.
 } \end{figure}   %%%%%%%%%%%%%%%%%%%%%%%%%%%%%%%%%%%%%%%%%%
\begin{figure}[htb]   % 2 %%%%%%%%%%%%%%%%%%%%%%%%%%%%%%%%%%%%%%
\centering     % \vspace{8mm}
 \includegraphics[scale=.40]{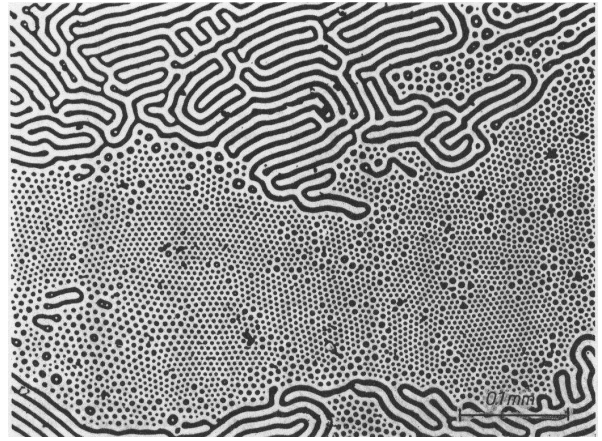}    % .42
%               %\vspace{-25mm}
\caption{\label{fig2}
Intermediate state of a type-I superconductor. The normal
domains are dark. Coexistence of triangular lattice of flux
tubes and of laminar domains. Tantalum disk of tickness
$d=33$ $\mu$$m$, diameter $D = 5$ mm, $T=1.2$ K , $B_a = 34$ mT.
(Courtesy U.\ Essmann). Optical microscope.
 } \end{figure}   %%%%%%%%%%%%%%%%%%%%%%%%%%%%%%%%%%%%%%%%%%
\begin{figure}[htb]   % 3 %%%%%%%%%%%%%%%%%%%%%%%%%%%%%%%%%%%%%%
\centering     % \vspace{8mm}
 \includegraphics[scale=.66,clip]{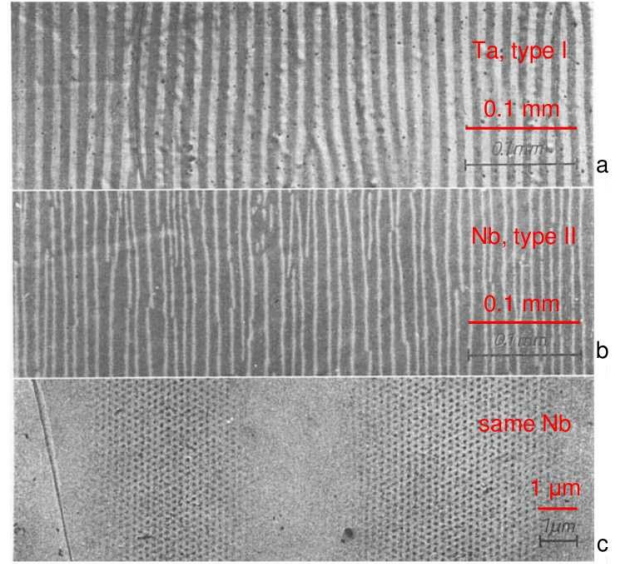}   % .68
%               %\vspace{-25mm}
\caption{\label{fig3}
Laminar domains in the intermediate state of pure superconductors
at $T = 1.2$ K. a) Type-I superconductor, Tantalum disc,
$d = 33~\mu$m, $D = 4$ mm, $B_a = 68$ mT, angle between magnetic
field and disc $15^\circ$. Optical microscope. b) Type-II superconductor,
Niobium disc, $d = 40~\mu$m, $D = 4$ mm, $B_a = 74$ mT. Optical
microscope. c) As b) but electron microscope. From [11].
 } \end{figure}   %%%%%%%%%%%%%%%%%%%%%%%%%%%%%%%%%%%%%%%%%%
 \begin{figure}[htb]   % 4 %%%%%%%%%%%%%%%%%%%%%%%%%%%%%%%%%%%%%%
 \centering
 \includegraphics[scale=.43]{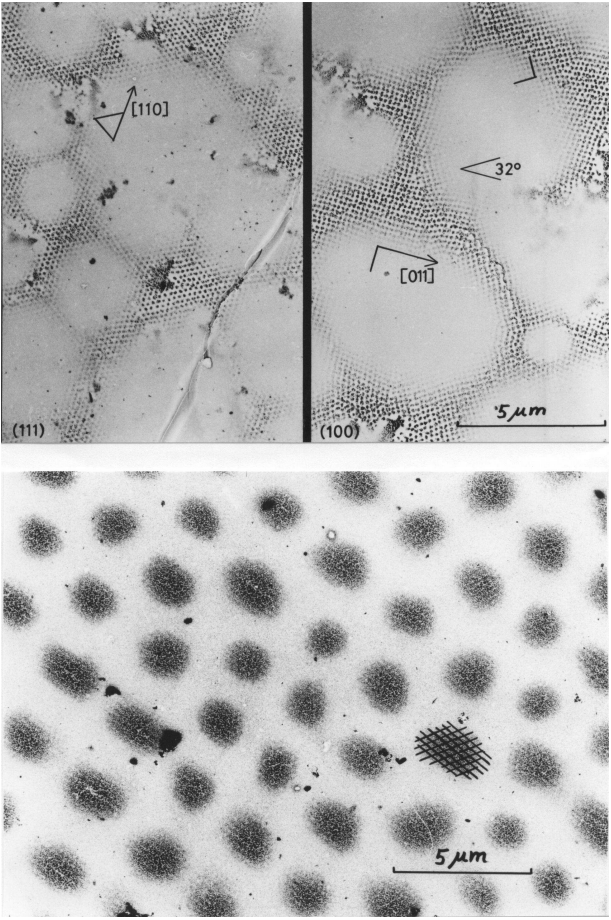}
\caption{\label{fig4}
 Flux-line lattice observed at the surface of
type-II superconductors in an electron microscope after decoration
with Fe microcrystallites (``magnetic smoke'').
{\it Top:~} High-purity Nb disks 1 mm thick, 4 mm diameter,
of different crystallographic
orientations [110] and [011], at $T=1.2$ K and $B_a = 800$ Gauss
($B_{c1} = 1400$ Gauss). Due to demagnetizing effects and the small
$\kappa \approx 0.70$, round islands of Meissner phase are
surrounded by a regular vortex lattice (``intermediate mixed state'').
{\it Bottom:~} High-purity Nb foil 0.16 mm thick at $T=1.2$ K and
$B_a = 173$ Gauss. Round islands of vortex lattice embedded in
a Meissner phase. (Courtesy U.\  Essmann)
 } \end{figure}   %%%%%%%%%%%%%%%%%%%%%%%%%%%%%%%%%%%%%%%%%%
 \begin{figure}[htb]   % 5 %%%%%%%%%%%%%%%%%%%%%%%%%%%%%%%%%%%%%%
 \centering
 \includegraphics[scale=.40]{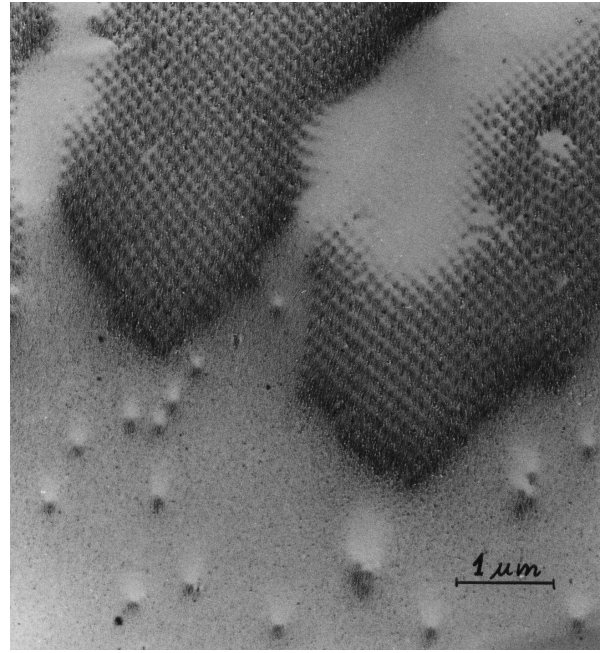}    %
\caption{\label{fig5}
Decoration of a square disk $5\times 5\times 1$ mm$^3$ of
high-purity polycrystalline Nb at $T=1.2$ K and $B_a = 1100$ Gauss.
As $B_a$ is increased, magnetic flux penetrates from the edges
in form of fingers which are composed of vortex lattice shown
enlarged in the right picture. The rectangular cross
section of the disk causes an edge barrier (section 8.5.).
 As soon as this  edge barrier is overcome, single flux lines or
droplets of vortex lattice (lower right) pull apart from these
fingers and jump to the center, filling the disk with flux from
the center. (Courtesy U.\  Essmann)
 } \end{figure}   %%%%%%%%%%%%%%%%%%%%%%%%%%%%%%%%%%%%%%%%%%
 \begin{figure}[htb]   % 6 %%%%%%%%%%%%%%%%%%%%%%%%%%%%%%%%%%%%%%
 \centering
 \includegraphics[scale=.42]{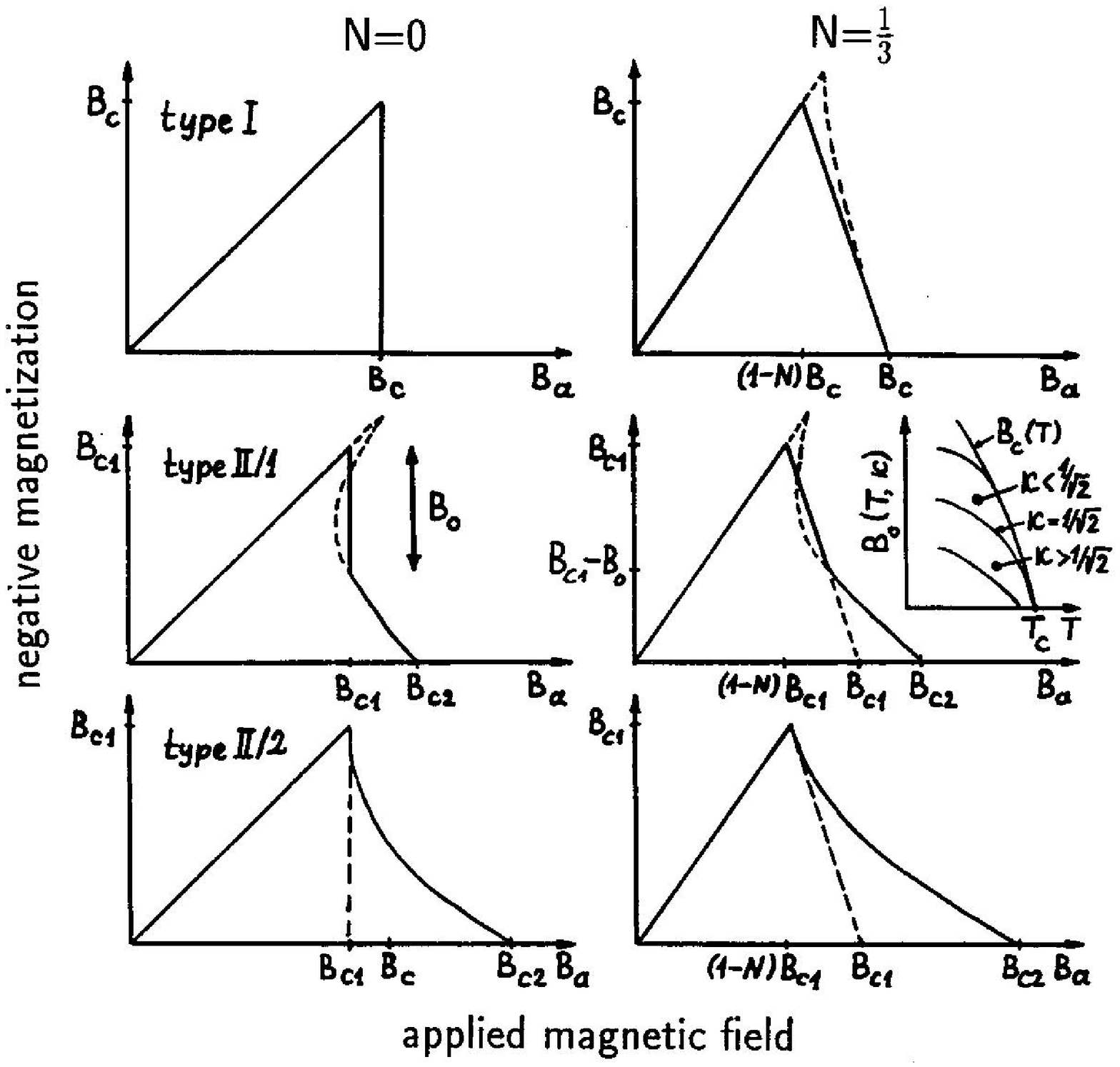}    %
\caption{\label{fig6} Ideal magnetization curves of long cylinders
or slabs in a parallel magnetic field $B_a$ (demagnetization
factor $N=0$, {\it left}) and of a sphere ($N=1/3$, {\it right}).
{\it Top row:~} Type-I superconductor with Meissner state for $B_a
<(1-N)B_c$, intermediate state for $(1-N)B_c \le B_a \le B_c$, and
normal state for $B_a>B_c$. The dashed line indicates the
enhancement of the penetration field when the wall energy of the
normal and superconducting domains in the intermediate state is
accounted for. {\it Middle row:~} Type-II/1 superconductors with
Ginzburg-Landau parameter $\kappa$ close to $1/\sqrt2$, for $N=0$
exhibit a jump of height $B_0$ in their induction and
magnetization at $B_a =B_{c1}$ due to an attractive interaction
between flux lines. For $N>0$, this jump is stretched over a
finite range of $B_a$, allowing one to observe an intermediate
mixed state with domains of Meissner phase and vortex-lattice
phase. {\it Bottom row:~} Type-II/2 superconductors with $\kappa
\gg 1/\sqrt2$. Meissner phase for $B_a < (1-N)B_{c1}$, mixed state
(vortex lattice) for $(1-N)B_{c1} \le B_a \le B_{c2}$, and normal
state for $B_a > B_{c2}$.
 } \end{figure}   %%%%%%%%%%%%%%%%%%%%%%%%%%%%%%%%%%%%%%%%%%

  Within the original GL theory in bulk
superconductors, with increasing applied magnetic field
$B_a = \mu_0 H_a$ the superconductor is first in the Meissner
state which has induction $B=0$. When $B_a$ reaches the
lower critical field $B_{c1}$ Abrikosov vortices start to
penetrate. With further increasing $B_a$ the vortex lines form
a more and more dense triangular lattice with induction
$B= (2/\sqrt3)\Phi_0 / a^2$ ($a$ = vortex spacing).
When $B_a$ and $B$ reach the upper critical field
$B_{c2} = \mu_0 H_{c2} = \Phi_0 /(2\pi \xi^2)$, the
order parameter vanishes and the superconductor
becomes normal conducting. Since $B(B_a)$ increases
monotonically (see Fig.~1 (top)), one may say that the
pressure of the vortex lattice is positive and the vortices
repel each other, held together by the applied magnetic field
$B_a$, which acts as an external pressure. From thermodynamics
follows that in equilibrium (in absence of pinning and surface
barriers) one has $B_a =\mu_0 H_a =\mu_0 \partial F/\partial B$
where $F(B,\kappa)$ is the GL free energy density. The numerical
GL solution $B_a(B)$ of the vortex lattice \cite{3}  for
$1/\sqrt 2 \le \kappa \le 10$ may be fitted well by the expression
(with $b=B/B_{c2}$, $h=B_a/B_{c2}$, $m=b-h= M/B_{c2}$,
$h_{c1} =B_{c1} / B_{c2}$):
 \begin{eqnarray}  % 1
   h(b) &=&  h_{c1} + {c_1 b^3 \over 1 +c_2 b + c_3 b^2 } \,,
                                \nonumber \\ \nonumber
   h_{c1} &=& { \ln \kappa +
   \alpha(\kappa) \over 2\kappa^2 }, ~~
   \alpha(\kappa) ={1\over 2} +
     {1+\ln 2 \over 2\kappa\! -\!\sqrt2\!+\!2} ,     \\
   c_1 &=& (1-h_{c1})^3 / (h_{c1}-p)^2  ,  \nonumber \\
   c_2 &=& (1-3h_{c1} +2p) / (h_{c1}-p) \,,\nonumber \\
   c_3 &=& 1 +(1-h_{c1})(1\!-2h_{c1}\!+p)/(h_{c1}\!-p)^2 ,
                                            \nonumber \\
   p &=&[\,(2\kappa^2 \!-1) \beta_A +1\,]^{-1}\,
  \end{eqnarray}
with $\beta_A = 1.15960$ (1.18034) for the triangular (square)
vortex lattice. At $\kappa=1/\sqrt2$ the $B(B_a)$ is a vertical
straight line at $B_a=B_{c1}=B_{c2}$ like in type-I
superconductors. This means the vortices do not interact with
each other and all vortex arrangements have the same free energy.
when $\kappa=1/\sqrt2$.

The vanishing vortex interaction for $\kappa=1/\sqrt2$ can also
be seen from Kramer's \cite{4} asymptotic interaction between two
vortices at large distance $r \gg \lambda$,
$V(r) = d_1(\kappa)K_0(r/\lambda) -d_2(\kappa)K_0(\sqrt2 r/\xi)$
where the constants are $d_1 =d_2$ at $\kappa=1/\sqrt2$.
A similar vortex potential \cite{5} was derived  such that it
reproduces the nonlocal elastic properties of the triangular
vortex lattice \cite{6} in the entire ranges of $B$ and $\kappa$
and for all wavelengths of the elastic strain. For parallel
vortices this approximate interaction potential reads
 \begin{eqnarray}  % 2
  V(r) = {\Phi_0^2 \over 2\pi \lambda'^2 \mu_0}
    [\, K_0(r/\lambda') - K_0(r/\xi') \,]
 \end{eqnarray}
with the lengths $\lambda' =\lambda/\langle |\psi|^2 \rangle^{1/2}
 \approx \lambda /\sqrt{1-b}$, $\xi' = \xi/\sqrt{2(1-b)}$
and $K_0(x)$ a modified Bessel function with the limits
$-\ln x$ ($x\ll 1$) and $(\pi/2x)^{1/2} \exp(-x)$ ($x\gg 1$).
Physically, the first (repulsive) term in $V(r)$ Eq.~(2) describes
the magnetic interaction and the second (attractive) term
originates from the overlap of the vortex cores that reduces
the condensation energy $\propto 1 -|\psi|^2$. Note that this
$V(r)$ has a finite maximum $V(0)$, approximately equal to half
the self energy of a vortex with flux $2\Phi_0$, since the two
terms diverging as $ \pm \ln r$ compensate.

  For $\kappa > 1/\sqrt2$ the potential (2) is repulsive.
At $\kappa = 1/\sqrt2$ it vanishes at all distances, which means
that all vortex arrangements have the same energy. This is an
exact result of the GL theory \cite{7}. The particular case
$\kappa = 1/\sqrt2$ and its surroundings ($\kappa \approx 0.71$,
$T \approx T_c$)  were investigated in detail in \cite{8,9}.

\section{2. Type-I Superconductors}    % 2

Superconductors with $\kappa < 1/\sqrt2$ are called as type I.
For $B_a <B_c=\mu_0 H_c$ (thermodynamic critical field)
these should exhibit ideal Meissner effect
with $B=0$ (except for a surface layer of thickness $\lambda$)
and $B =B_a$ for $B_a >B_c$. But in samples of finite size the
demagnetization factor $N>0$ allows the observation of an
``intermediate state'' containing superconducting and normal domains
with induction $B=0$ and $B=B_c$, respectively, in the range
$(1-N)B_c < B_a < B_c$ \cite{10}. Figure 2 shows a decoration
picture of the type-I superconductor Tantalum, a disk with
thickness $d=33~\mu$m, diameter $D=5$ mm, at $T=1.2$ K and
$B_a =34$ mT, taken in 1969, from the review \cite{11}.
The picture shows the coexistence of triangular lattice of flux
tubes and of laminar domains. As recently shown by
Prozorov \cite{12,13}, the equilibrium topology of the
intermediate state are flux tubes (in absence of bulk pinning
and geometric barrier). Interestingly, pin-free disks and strips
of constant thickness show hysteretic magnetization curves
due to edge barriers for penetrating flux tubes and exiting
lamellae. This topological hysteresis of type-I superconductors
was observed and calculated analytically already in
1974 \cite{14}, where it was shown that ellipsoidal samples
exhibit no magnetic hysteresis. Magnetic hysteresis loops in
pin-free type-I superconductor disk and strips look very similar
as the irreversible magnetization loops in pin-free type-II
superconductors with edge barrier \cite{15},
in particular, they go through the origin
($M=B-B_a=0$ at $B_a=0$, no flux is trapped) and become
reversible at large $B_a$ (at $B_a > B_{c1}/2$ for type-II).

  The structure of the flux tubes or lamellae in type-I
superconductors and their widening (mushrooming) and branching
as they approach the surface were calculated, e.g., in
\cite{16,17}. In thin films of type-I superconductors a
vortex lattice occurs though $\kappa < 0.71$ \cite{18}.
In particular, for $\kappa = 0.5$ the shear modulus $c_{66}$
of this vortex lattice is positive at all inductions
$0 < B < B_{c2}$ when the film thickness is $d \le \xi$,
see Figs.~9 and 13 in \cite{19}, and the magnetization curve
$-m(h) = -m(b)$ (Fig.~4 in \cite{19}) has positive curvature
in this case. These GL results show that in thin (and infinitely
large) films in perpendicular field $B_a = B$ the vortices
may repel each other even when $\kappa < 0.71$ since the
long-range repulsive interaction via the magnetic stray field
dominates \cite{20}. For the vortex interaction in thin films
of finite size see \cite{21}.

\section{3. Attractive Vortex Interaction}   % 3

  The first successful decoration experiments evaporated an iron
wire in a 1 Torr He atmosphere that produced ``magnetic smoke''
which settles on the surface of the superconductor and marks
the ends of the flux tubes or flux lamellae (Fig.~2,3a) or of the
vortices (Figs.~3c,4,5) \cite{22}. In the type-II superconductors
Vanadium and Niobium (see Fig.~3 of \cite{11}) and PbIn alloys
they showed more or less regular triangular vortex lattice with
structural defects, e.g., dislocations and vacancies. These
pictures were consistent with the prediction from GL theory
that the vortices repel each other. However, later decoration
of very pure Nb showed a domain structure similar to that
observed in type-I superconductors but with the normal domains
replaced by domains of vortex lattice. The Figs.~3a and 3b
compare the laminar domains in the type-I superconductor
Tantalum and in the type-II superconductor Nb. In an optical
microscope these domains {\em look very similar}, but the
electron microscope (Fig.~3c) reveals that in Nb the domains
which carry the magnetic flux consist of vortex lattice with
constant lattice spacing $a_0$.

  In Fig.~4 (top) one sees islands of Meissner state embedded
in a vortex lattice, while Fig.~4 (bottom) shows Meissner state
with islands of vortex lattice. In Fig.~5 tongues of vortex
lattice penetrate from the specimen edge into the Meissner
state of high-purity polycrystalline Nb.

  This observation of an upper limit $a_0$ of the vortex
spacing in the intermediate state of low-$\kappa$ type-II
superconductors, and its explanation by an attractive
interaction, suggests that the magnetization curves in the
ideal (pin-free) case should exhibit a jump of height
$B_0 = (2/\sqrt3)\Phi_0/a_0^2$ at the field of first vortex
penetration. This means, as soon as it is energetically
favorable for vortices to penetrate, they immediately jump
to their equilibrium distance $a_0$, which for $N=0$ means
a uniform induction $B_0$, see Fig.~1 (bottom). Such jumps of
$B(B_a)$ from 0 to $B_0$ were indeed observed in clean Nb
and in lead alloys with increasing content of Thallium
(which changes $\kappa$), and excellent agreement of $B_0$
with the directly measured $a_0$ was observed \cite{23,24,25}.
Jumps $B_0$, and thus vortex attraction, were also
observed in the TaN system with varying nitrogen content
and various $\kappa = 0.35$ to $\kappa = 1.05$ \cite{26}.
See also the reviews \cite{11,27}.

  The jump $B_0$ in the ideal magnetization curve $B(B_a)$
was confirmed by computations \cite{28} based on the
microscopic BCS-Gor'kov theory that reduces to the GL theory
in the limit $T \to T_c$. Earlier computations \cite{29}
confirming the vortex attraction were based on the extended
GL theory of Neumann and Tewordt \cite{30} that keeps all
correction terms to lowest order in $1- T/T_c$.
The BCS-Gor'kov computations \cite{28} for clean
superconductors yield an S-shaped curve $B(B_a)$ with an
unstable part (dashed curve in Fig.~1 (bottom)) that has to be
replaced by a vertical line of height $B_0$ that cuts the area
under the dashed line in two equal parts (Maxwell construction).
Useful interpolation formulae for this S-shape are given
in \cite{31}.

  The phase diagram of low-$\kappa$ clean superconductors in
the $\kappa-t$ plane ($t=T/T_c$) was measured by Auer and
Ullmaier \cite{26} and computed from BCS-Gor'kov-Eilenberger
theory by Klein \cite{32}; this is discussed in \cite{33}.
Near $\kappa = 0.71$ with increasing $t$, this phase diagram
shows the transition from type-I to usual type-II
(termed type-II/2) via a region where the vortices attract
each other, termed type-II/1 superconductors.
Figure 6 shows the magnetization curves $-M(B_a)$, $M=B-B_a$,
for type-I, II/1, and II/2 superconductors shaped as cylinders
($N=0$, left column) and spheres ($N=1/3$, right column),
schematic from \cite{27}. Recently by neutron-scattering
a kind of phase diagram of vortex lattices and normal
and intermediate-mixed states has been obtained
for pure Nb cylinders of 14 mm length and 4 mm diameter \cite{34}.
So, vortex attraction in pure Niobium is still an active field.

  The above attractive interaction originates from BCS-corrections
to the GL theory (which, strictly spoken, is valid only at $T=T_c$)
and at $0 < T < T_c$ occurs for $0.71 \le \kappa \le 1.5$ \cite{28}.
The London theory that follows from GL theory when
$\kappa^2 \gg 1$ and $b \ll 1$, has purely repulsive vortex
interaction, namely, the first term in Eq.~(2). However,
many corrections to the simple GL or London theories are
expected to modify the monotonically decreasing interaction
potential at large distances, $V(r) \propto \exp(-r/\lambda)$,
such that $\lambda$ becomes complex. This, in principle, causes
an oscillating potential, whose first minimum may occur at
large distances where the amplitude of the potential is small.

  Such corrections may have various origins, e.g.,
the spatially varying order parameter may lead to non-local
electrodynamics \cite{35}. Linear nonlocal electrodynamics
relates the supercurrent density ${\bf j}$ to the vector
potential ${\bf A}$ (in Fourier space and London gauge) by a
kernel $Q(k)$, ${\bf j(k)} = -Q(k) {\bf A(k)}$. The simple
``local'' London theory has $Q(k) = Q(0) = \lambda^{-2}$.
An elegant non-local
extension of London theory uses the Pippard kernel
$Q_P(k)$ \cite{36}, which introduces a Pippard length $\xi_P$.
This Pippard theory leads to a vortex field $H(r)$ and vortex
potential $V(r) = H(r) \Phi_0$ that oscillates at large
distances (``field reversal'' \cite{37,38}). The corresponding
BCS kernel $Q_{BCS}(k)$ is an infinite sum of such Pippard
kernels and its vortex field $H(r)$ and potential $V(r)$
exhibit similar oscillations \cite{37}, see also \cite{38}.
We note that some of the early theories quoted in \cite{26}
were flawed.

In anisotropic superconductors vortex attraction may even occur
in the linear London theory, which now has different penetration
depths $\lambda_a$, $\lambda_b$, and $\lambda_c$ for supercurrents
flowing along the crystalline a, b, and c axes \cite{27}. A field
reversal and vortex attraction may then occur when the vortices
are at an oblique angle \cite{39,40,41}.

In layered superconductors an anisotropic pair potential between the
pancake vortices is investigated by  Feigel'man et al. \cite{42}
and Clem \cite{43}. This potential has an intra-layer logarithmic
repulsive part and an inter-layer attractive part, both arising
from the magnetic interaction. This potential is used by
Jackson and Das \cite{44,45} for the study of freezing of
vortex fluids by a density functional method, yielding a nearly
first order transition from a fluid to a triangular
Abrikosov vortex lattice.
Brandt et al.\ \cite{46} predicted long-ranged fluctuation-induced
attraction of vortices to the surface in layered superconductors
by a type of Casimir force. This work inspired Blatter and
Geshkenbein \cite{47} to study long-range van der Waals type
inter-vortex attraction in the anisotropic layered materials,
which arises from the thermal fluctuations of vortices.
At low magnetic fields, the inter-vortex separation
being very large, even weak attraction influences the phase
diagram near the $H_{c1}$ boundary. While the above forces are of
thermal origin, van der Waals attraction mediated by impurities
is studied by Mukherji and Nattermann \cite{48}.

  It is conceivable that also a two-component GL
theory may lead to an oscillating vortex-vortex
interaction with attractive tail and more or less pronounced
minimum if its input parameters are chosen appropriately.
Such a potential was calculated in \cite{49,50}. In the same
work some of the decoration \cite{49} and scanning SQUID
microscopy \cite{50} images taken
during field cooling of MgB$_2$ platelets suggest that at
the (unknown) temperature where the depicted vortex positions
during field cooling were frozen, the vortex interaction had
a minimum that determined the dominating vortex distance in
the observed vortex chains and clusters.

While Refs.~\cite{49,50} do not mention any of the above
listed other cases of vortex attraction, the
supplement \cite{51} refers to previously observed and
well understood vortex attraction in pure Nb. Note that
possible observation of vortex attraction in MgB$_2$
does not mean that the input parameters and interaction
potential of simulations that lead to similar vortex
arrangements are true or unique.
In the following the case of the 2-band superconductor
MgB$_2$ will be discussed in some detail.

\section{4. BEYOND THE STANDARD BCS MODEL}  % 4

The BCS theory is the standard model for understanding various
basic properties of conventional metallic superconductors. In
this simple theory the Fermi surface is isotropic. The attractive
electron-phonon interaction is a constant in a small shell around
the FS. In 1958 Gorkov showed the equivalence of the BCS theory
with the one-order parameter isotropic Ginzburg-Landau theory.
The latter is a powerful phenomenological theory suitable for
practical applications. On the other hand a more rigorous
approach is the Eliashberg theory suitable for strong coupling
of electron-phonon interaction.

In the past three decades a number
of novel superconductors have been discovered with many exotic
properties beginning with the high-$T_c$ cuprates. Others are
doped fullerides, ruthenides, nickel borocarbides, magnesium
diborides and pnictides etc. Basic understanding of
superconductors requires the correct information about its
electronic and vibrational structures and electron-phonon
interaction. Fortunately an ab-initio theory like the density
functional theory helps us to meet this type of requirement.
By using a self-consistent elctron potential arising out of
ionic, Hartree, exchange and correlations, one computes the
quasi-particle dispersion throughout the Brilliouin zone and
obtains the Fermi surface.

In a superconductor the formation of Cooper pairs depends
heavily on the geometry of the Fermi surfaces. Then comes the
electron-phonon interaction that mediates the pairing. In view
of the complicated band structures and their Fermi surfaces,
the novel superconductors are treated beyond the standard
BCS model. We are reminded that single-sheeted Fermi surfaces
only come from the simple metals like alkalies considered
in the jellium model and those are not superconductors under
ambient conditions. Therefore, if the multi-sheeted Fermi
surfaces are connected in the entire Brilliouin zone, then an
anisotropic GL theory is useful. Otherwise for unconnected
Fermi surfaces one can use either (i) a multi-band BCS
theory \cite{52} or (ii) a generalization of the GL theory
invoking many (or many components of) order parameters \cite{53}.
This two-component order-parameter theory was used to explain
successfully two specific-heat transitions in the high $T_c$
superconductor YBCO (see Choy et al.\ in \cite{53}).

As said before, in reality the bands are formed due to ionic and
electronic self-consistent potential. In principle we are dealing
with a charged (Coulomb) system, hence the bands are not entirely
independent, and further due to the electron-phonon interaction
(whether small or large) the bands are coupled. For this reason
multi- (inter- and intra-)band interaction terms explicitly
appear in the microscopic Hamiltonian \cite{52}. In the case of
multi-order parameter GL theory, one has the superconducting free
energy composed of individual parts of the GL free energy and the
additional free energy due to coupling between the order
parameters as mentioned above. This choice of coupling is
non-unique. If $\psi_1$ and $\psi_2$  are two order parameters the
coupling or interaction term can be one or more terms from the
following list:
$$ \gamma_1 |\psi_1|^2 |\psi_2|^2 , ~
   \gamma_2 (\Pi^* \psi^+_1 \, \Pi \psi_2 + {\rm hc}) , ~
   \gamma_3 (\psi^+_1 \psi_2 + \psi_1 \psi^+_2 )
$$
                   %\end{document}
with $\Pi = \nabla + 2\pi i {\bf A} / \Phi_0$.
The first term of this list is a systematic expansion of GL,
the second term means coupling through the gradient of
order parameter and vector potential {\bf A}, and the third term
is the internal Josephson coupling term. This last term is
supposed to provide a ``minimal coupling'' \cite{54,55,56}. Instead
of the phenomenological GL theory, we start with the microscopic
Bogolyubov pairing Hamiltonian and use the mean-field approach;
we derive the GL functional with a coupling term that will be
Josephson-like. An important point here is that
the coefficients $\alpha_i$ of $|\psi_i|^2$, and the
coefficient  $\gamma_3$  of the Josephson term, do depend both
on the intra-band and inter-band pairing. Therefore, as a matter
of principle the {\em order parameters can never be
independent}. There is a huge amount of literature on both
microscopic and GL theory for the multi-band cases. We do not
intend to review that here, see for example\cite{55}. We shall
return to the interacting bands in the next Sections.

In the earlier sections
we have discussed the physics of vortex matter with
particular emphasis on the value of $\kappa$ and the type of
superconductivity (type-I or type-II). We also discussed the
behavior of pure Nb at temperatures below the
GL regime.  In the conventional single gap superconductor
within GL theory there are a well defined
coherence length $\xi$ and penetration depth $\lambda$. In terms
of them one defines $\kappa = \lambda/\xi$. As stated earlier
its unique value, if less or bigger than 0.71, defines precisely
type-I or type-II superconductors, in which the surface
free energy between normal and Meissner domains is positive
or negative, respectively. This definition tells about the
equilibrium states of a superconductor.
Here with this brief background we shall discuss the interesting
physics of the novel superconductor MgB$_2$ and the vortex matter
therein. Needless to say that it has become a news item in the
past one year with the claim that a new form of matter is
discovered \cite{49,50}. Detailed discussions follow in Sec.~6.

\section{5. VORTICES IN MgB$_2$}  % 5

In 2001 by a surprising discovery Nagamatsu et al. \cite{57}
reported $T_c$ of an intermetallic MgB$_2$ around 39 K.
Soon after (see \cite{58} database for the first three years)
it was realized that MgB$_2$ offers a host
of novel properties that has been explored during the past
9 years. Magnesium diboride has a crystallographic structure
consisting of honeycomb boron layers separated by magnesium
layers, which have also honeycomb structure. Magnesium atoms
are ionized in this structure donating their s-electrons to the
conduction band. There are two 3-dimensional $\pi$ bands arising
from the boron $p_z$ orbitals. One of them is electron-like,
whereas the other is hole-like. From the $p_{x,y}$ orbitals
two $\sigma$ bands occur. They are 2-dimensional confined to
the boron planes. Bonds within the boron layers are strongly
covalent, whereas between the layers they are metallic. From
this a complicated Fermi surfaces geometry emerges. Two
cylinders around the $\Gamma$-A-$\Gamma$ lines are two $\sigma$
bands, and two webbed tunnels are attributed to $\pi$ bands,
see Fig.~7 (from Mazin and Antropov \cite{59}).
This overall picture is consistent
with all the electronic structure calculations \cite{59}.

 \begin{figure}[htb]   % 7 %%%%%%%%%%%%%%%%%%%%%%%%%%%%%%%%%%%%%%
 \centering    \vspace{-5mm}
 \includegraphics[scale=.46,clip]{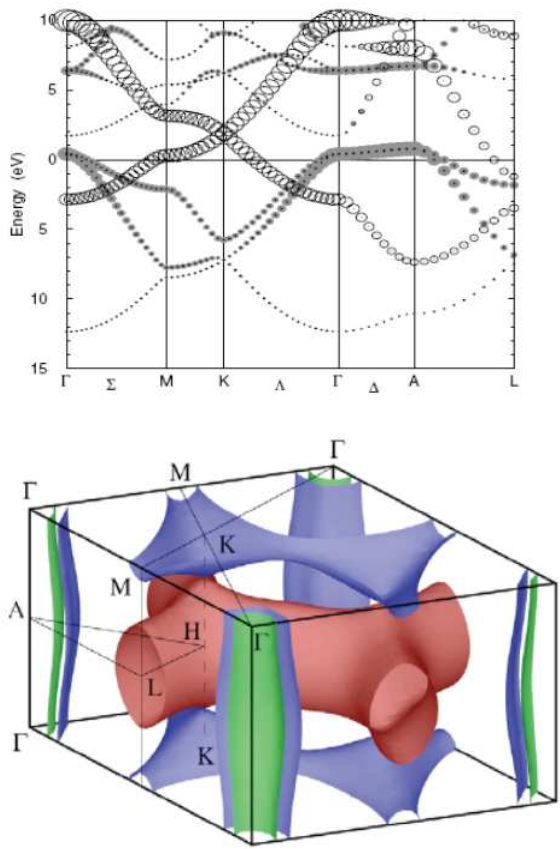}    % .37
    \vspace{-10mm}
\caption{\label{fig7}
 Band structure and Fermi surface of MgB$_2$ from Mazin and
Antropov \cite{59}. Green and blue cylinders (hole-like) are
from the $\sigma$ bands, and the blue (hole-like) and red
(electron-like) tubular networks are from the $\pi$ bands.
 } \end{figure}   %%%%%%%%%%%%%%%%%%%%%%%%%%%%%%%%%%%%%%%%%%

It is further found \cite{60} that the optical bond-stretching
$E_{2g}$ phonons couple strongly to the holes at the top of
$\sigma$ bands, whereas the 3-dimensional $\pi$ electrons are
weakly coupled to the phonons. For this reason of different
coupling strengths of $\sigma$ and $\pi$ bands, a two-gap
superconductor merges. The $E_{2g}$ phonons strongly coupled
with $\sigma$ band, produce high $T_c$ in MgB$_2$. A further
coupling with the $\pi$ bands enhances $T_c$.
Table 1 presents values of intra- and inter-band (electron-phonon)
coupling-constant matrix-elements calculated by various groups.
The numbers show that the electron-phonon coupling $\lambda$
(a dimensionless parameter \cite{61} not to be confused with the
penetration depth $\lambda$) for the $\sigma$ band is very
strong and for the $\pi$ band it is weak.
The inter-band coupling is still weaker but appreciable. These
numbers are useful ingredients to be incorporated in the
two-order parameter calculations.
In a first principles calculation using Eliashberg theory,
Geerk et al.~\cite{60} emphasize the importance of
inter-band pairing and low energy optical modes arising out of
boron atoms vibrating perpendicular to the boron planes.
\\\\
\begin{tabular}{c c c c l} \hline
 $\lambda_{\sigma\sigma}$ & $\lambda_{\pi\pi}$    &
 $\lambda_{\sigma\pi}$    & $\lambda_{\pi\sigma}$ & \\[1mm]
                                              \hline\hline
 1.017~~ & 0.448~~ & 0.213~~ & 0.155~~ & ~Golubov et al.~[60]
                                                    \\[1.5mm]
 0.96 &  0.29  & 0.23  & 0.17   & ~Liu et al.~[60]  \\[1.5mm]
 0.78 &  0.21  & 0.15  & 0.11   & ~Choi et al.~[59] \\[1mm]
                                                    \hline
\end{tabular}
\\[3mm]
Table 1. Matrix elements of electron-phonon interactions
\\

In the past nine years a variety of experiments have shown that
MgB$_2$ is a two band superconductor. The main experiments are
specific heat (Choi et al. in \cite{59,62}, point contact and
tunnelling spectroscopy \cite{63,64}, Raman spectroscopy \cite{65},
penetration depth measurements \cite{66}, angle-resolved
photo-emission spectroscopy \cite{67}, small-angle
neutron-scattering \cite{68}, and muon-spin relaxation
studies \cite{69} etc. The value of the gap on
the $\sigma$ bands $\Delta_\sigma$ ranges from 6.4 to 7.2 meV,
while on the $\pi$ bands $\Delta_\pi$ varies from 1.2 to 3.7 meV.
Almost all parameters such as coherence lengths ($\xi_\sigma$ and
$\xi_\pi$) \cite{66}, upper critical fields ($H_{c2}(\sigma)$ and
$H_{c2}(\pi)$) \cite{70,71}, and penetration depths ($\lambda_\sigma$
and $\lambda_\pi$) \cite{66} are anisotropic. Besides this the
anisotropic factor is also strongly temperature dependent.
These experiments convincingly suggest that MgB$_2$ is a
two band superconductor, where phonons mediate pairing
in the respective bands. Further, it is investigated that
inter-band coupling helps to increase $T_c$. In the absence of
inter-band coupling, it may give two isolated superconductors
with two lower $T_c$.

\section{6. A NEW PHASE OF MgB$_2$ !}  % 6

 In a recent paper Babaev and Speight \cite{72} studied
phenomenologically a two-band superconductor with two
order parameters, starting with a free energy that includes a
Josephson-like interaction $\gamma_3$ between the two bands.
Choosing $\gamma_3 \to 0$ in the decoupled band limit,
they considered $\kappa_i$ ($i=1,2$) in two different regimes
to produce type-I ($\kappa_1 < 0.71$) and type-II
($\kappa_1 < 0.71$) materials. Their prediction leads to,
what they call a ``semi-Meissner state'', particularly in the
case of very low magnetic fields. Babaev and Speight claim
that this state is thermodynamically stable. Instead of
homogeneous distribution, the vortices form aperiodic clusters
or vortexless Meissner domains, arising out of short range
repulsion and long range attraction. Below we shall argue
that in the particular case of MgB$_2$, this is an
unrealistic construction.

In what is claimed as a remarkable manifestation of a new type
of superconductivity in MgB$_2$, Moshchalkov and
co-workers \cite{49,50} reported last year a novel
``type-1.5 superconductor'', contrary to type-I and type-II
superconductors. It appears these authors are influenced by
the idea of ``semi-Meissner state'' of Babaev and Speight.
Apparently both names carry similar meaning. Let us analyse
carefully the results of Moshchalkov et al.

Similar to Babaev and Speight, Moshchalkov et al.\ use a two-band
GL functional with an inter-band coupling term. They constructed
the GL parameters $\kappa_i = \lambda_i/\xi_i$ ($i$ stands for
$\sigma$ and $\pi$ bands of MgB$_2$) from the measured values of
energy gaps, Fermi velocities and plasma frequencies of
respective bands. The values are so chosen that
$\kappa_\sigma > 0.71$ (=3.68) and $\kappa_\pi < 0.71$ (= 0.66).
Thus, MgB$_2$ has properties of both type-I and type-II
{\em simultaneously}, unlike in the case of Nb where
$\kappa$ is close to 0.71 as discussed above. This construction
is in absence of the coupling term, so the bands are fully
independent/noninteracting. We have stated earlier that energy
gaps $\Delta_\sigma$ range from 6.4 to 7.2 meV, while the
$\Delta_\pi$ varies from 1.2 to 3.7 meV. Moshchalkov et al.\ chose
the value of $\Delta_\sigma = 7.1$ meV and $\Delta_\pi = 2.2$ meV
from some experiments and computed $\xi_i$ from the one-band
BCS relation $\xi_i = h v_F /(2 \pi^2 \Delta_i)$. Then they used
calculated plasma frequencies of respective bands, from which
they obtained the London penetration depths
$\lambda_\sigma(0)= 48.8$ nm and $\lambda_\pi(0)= 33.6$ nm
and finally $\kappa_i = \lambda_i / \xi_i$.

Here we note a few points.

A. (1) The numerical estimates for $\xi_i$ are obtained by using
the one-band BCS formula. (2) One band calculated plasma frequency
is used to compute $\lambda_i$. One does not know how these
are relevant to interacting two-band MgB$_2$.

B. For critical fields Moshchalkov et al.\ gave their measured
values for $H_{c1}(0) = 241$ mT, $H_{c2}(0)= 5.1$ T for the
$\sigma$ band. From these data they also obtained $\lambda_\sigma$
and $\xi_\sigma$¦ from the one band formula. The value of
$H_{c1}(0) = 241$ mT measured for their sample is much larger
than the values reported by others, which are in the range of
25 - 48 mT, see Fig.~25 of Ref.~\cite{71}. Another point of
concern is the reported value of thermodynamic critical field
$H_c(0) = 230$ mT \cite{73}, whereas Moshchalkov et al.\
have $H_{c1}(0) = 241$ mT  higher than $H_c(0)$.

C. The $\lambda$ that Moshchalkov et al.\ estimated are smaller
by a large factor as can be verified from the Table 1 of
Golubov et al.~\cite{74}. This Table lists the values of
$\lambda(0)$  from a variety of experiments. Smaller
$\lambda$ can make the $\kappa$ smaller. This possibility has
made $\kappa_\pi < 0.71$. Another important point is that
using two $\lambda$ for a single flux line is questionable.
One has to solve a single equation for the magnetic field
(in two-order parameter GL/BCS theory) to obtain a
single $\lambda$, even in the decoupled band limit. One has
$\lambda \propto (\sum_i |\psi_i|^2 /m_i)^{-1/2}$,
containing a sum of two superfluid densities.

An often heard argument is that whether there is one $\kappa$
or two $\kappa$, there will be only one surface free energy
between Meissner state and normal state, which would be either
positive, zero or negative. Thus, superconductors are either
of type-I or type-II. However, this argument tacitly
assumes that boundaries occur only between Meissner and
normal states. But as figures of the intermediate-mixed
state show (see Figs.~3 to 5), there
the boundary occurs between Meissner and vortex
states. In these experiments the wall energy between
Meissner state and normal state is negative (leading to
a vortex state) but the wall energy between Meissner
state and the vortex state is positive, causing the
observed domain structure. So, there are indeed
experimental situations that exhibit features of both
type-I and type-II behavior. As stated in  Sec.~3,
one condition for this intermediate-mixed state to occur
is that the vortex interaction has an attractive tail,
or that the theoretical magnetization curve is S-shaped,
see Fig.~1 (bottom).
But in addition, the demagnetization factor of the
decorated specimen has to be $N > 0$, else only
Meissner state or a uniform vortex state will occur;
the detailed appearance further depends on pinning,
on the edge barrier, and on the magnetic history.
This intermediate mixed state has been known for long
time from experiments \cite{11,22,23,24,25,26,34,75}
and computations \cite{28,29,31,32,33} for pure Nb,
TaN, PbIn, and PbTl. In this respect, the decoration and
scanning SQUID results of \cite{49,50} are not so new.

As discussed above, several origins of vortex
attraction are conceivable, depending on the material,
its purity, magnetic history, and temperature.
Computer simulations using such a vortex interaction
(short-range repulsive, long-range attractive) yield
vortex arrangements that are similar to the observed
ones \cite{49,50}. The details of the resulting clusters
or chains also depend on the type and strength of vortex
pinning that is always present and should be included in
such simulations, in particular at very low induction $B$.

\section{7. Summary and Conclusions}  % 7

  Type-I superconductors with demagnetization factor
$N>0$ in the field range $(1-N)B_c < B_a < B_c$ are in the
intermediate state showing normal conducting lamellae or tubes
inside which $B=B_c$, embedded in Meissner state with $B=0$.
Type-II superconductors in the field range
$(1-N)B_{c1} < B_a < B_{c2}$ contain vortices which repel
each other (mixed state). Under several conditions the vortex
interaction may have an attractive tail, leading to observation
of vortex clusters or lamellae similar as in type-I
superconductors, but filled with vortices, and to vortex chains.
This ``intermediate-mixed state'' has been known since
forty years in pure Nb, TaN \cite{26}, PbIn, and
PbTl \cite{75} at temperatures
sufficiently low such that GL theory does not apply.
The recent possible observation of vortex attraction in
the two-band superconductor MgB$_2$, if one agrees with
this interpretation of the decoration \cite{49} and
scanning SQUID \cite{50} images, may be due to the
complexity of the required microscopic theory. But we
question if these images hint at ``a totally new state''
that ``behaves in an extremely unusual way'' \cite{49}.
It should also be considered that at the very low inductions
of these images even very weak pinning will influence the
vortex positions and may cause vortex clustering \cite{76}.

  MgB$_2$ is a ten year old new inter-metallic superconductor which
has many fascinating properties, particularly a high
$T_c\approx39$ K. We briefly discussed its electronic and
geometrical
structure in the pure case. For understanding the occurrence of
superconductivity we showed in Fig.~7 its two sets of $\sigma$
and $\pi$ bands. The electrons in these bands are interacting
via Coulomb and phonon interactions. There are unambiguous
calculations and experiments which suggest the high strengths of
 inter-band and comparable intra-band interactions in this
two-band material. It has been a consensus that MgB$_2$ is a
phonon-mediated superconductor. The higher $T_c$ is a result
of coupling between the two Fermi surfaces
(see e.g.\ Geerk et al.\ in \cite{60}). This information provides
a compelling reason to adopt a two-band model, either by
microscopic BCS-Eliashberg or GL theory (see e.g.\
\cite{52,53}, Zhitomirsky and Dao in \cite{55}, \cite{77},
and the review \cite{78}). Thus an argument of two independent
bands is never tenable for MgB$_2$ that exhibits one single
$T_c$. Hence no separate two superconducting regimes can occur
in MgB$_2$.

   A single vortex in a two-band superconductor will have a
complicated vortex core that results from solving
two coupled order parameter equations.
Contrary to the remark by Babaev and Speight \cite{60}, the core
size and shape depends nontrivially on the coupling of two
order parameters and also on temperature. The next issue is about
the penetration depth $\lambda$. This quantity comes out of the
GL equation for the  vector potential.
In the presence of electromagnetic field coupling with the
order parameter
[$\gamma_2 (\Pi^* \psi^+_1 \, \Pi \psi_2)$ + hc]
as mentioned earlier in Sec.~4, $\lambda$ depends on the
superfluid densities in a nontrivial fashion, see Eq.~9 and Fig.~3
of \cite{79}. In brief, mixing two components as visualized by
Babaev et al.\ and Moshchalkov et al., will not produce a new kind
of superconductivity, because in the mixture $\lambda$ is
determined self-consistently. Table 1 of Golubov et al.~\cite{74}
lists 12 values of the zero-temperature London penetration
depths $\lambda_L(0)$ from various experiments
with MgB$_2$; these values exceed
the reported values of the coherence lengths $\xi$, and thus
$\kappa \gg 0.71$. Therefore, type-I condition will be hardly
satisfied. For this reason it will be instructive to assess
the ``semi-Meissner'' or ``type-1.5'' conditions in relation to
Moshchalkov's experiments \cite{49,50}.

Another question is: Can
one have two different penetration depths $\lambda$ in the
same material and for the same direction of the supercurrent
density ${\bf j}$\,? Usually $\lambda$ is defined via the
general linear relation (valid for small $j$ and small $B$
in isotropic superconductors, see Sec.~3)
${\bf j(k)} = -Q(k) {\bf A(k)}$ by $Q(0) = \lambda^{-2}$
or $\lambda = Q(0)^{-1/2}$. In general one has
${\bf j} = \delta F / \delta {\bf A}$ where $F\{ {\bf j, A}\}$
is the free-energy functional of a given theory and ${\bf A}$
the vector potential. This definition allows only for one
single $\lambda$.

In conclusion, from our careful analysis we argue that the
hypothesis of ``type-1.5 superconductivity'' as put forward
in \cite{49,50} and in the abstract of Moshchalkov's talk
at this conference, is unrealistic and unfounded for MgB$_2$
that has one single transition temperature $T_c$.
Anyway, it cannot be precluded that the vortex-vortex
interaction, if calculated microscopically (e.g., by the
quasiclassical Eilenberger method \cite{31,32,80,81}) may
have an attractive tail due to the complexity of the problem.

%%%%%%%%%%%%%%%%%%%%%%%%%%%%%%%%%%%
%\acknowledgments

\section{Acknowledgments}  % 8

We thank U.~Essmann, R.\ P.\ Huebener, John R.\ Clem,
A.~Gurevich, R.~Prozorov, V.\ G.\ Kogan, and S.-P.\ Zhou
for helpful discussions.

%This work was supported by the German Israeli Research Grant
%Agreement (GIF) No G-901-232.7/2005.

{}

\end{document}